\documentclass{PoS}
\usepackage{booktabs}
\usepackage{multirow}

\title{Development of Radiation Hard Scintillators}

\ShortTitle{Development of Radiation Hard Scintillators}

\author{\speaker{Emrah Tiras}$^{1}$, {James Wetzel$^{1}$}, {Burak Bilki$^{1}$$^{,2}$}, {David Winn$^{3}$}, {Yasar Onel$^{1}$}\\
\\
$^{1}$University of Iowa, Iowa City, IA, USA\\
$^{2}$Beykent University, Istanbul, Turkey\\
$^{3}$Fairfield University, Fairfield, MD, USA\\
\\
E-mail: \email{emrah-tiras@uiowa.edu}}

\abstract{Modern high-energy physics experiments are in ever increasing need for radiation hard scintillators and detectors. In this regard, we have studied various radiation-hard scintillating materials such as Polyethylene Naphthalate (PEN), Polyethylene Terephthalate (PET), our prototype material Scintillator X (SX) and Eljen (EJ). Scintillation and transmission properties of these scintillators are studied using stimulated emission from a 334 nm wavelength UV laser with PMT before and after certain amount of radiation exposure. Recovery from radiation damage is studied over time. While the primary goal of this study is geared for LHC detector upgrades, these new technologies could easily be used for future experiments such as the FCC and ILC. Here we discuss the physics motivation, recent developments and laboratory measurements of these materials.}

\FullConference{38th International Conference on High Energy Physics\\
		3-10 August 2016\\
		Chicago, USA}

\begin{document}

\section{Introduction}
The increase in energy at particle physics experiments offers an excellent oppurtunity to disclose the properties of the exotic particles and discover new particles beyond the Standard Model. However, this results in more radiation and  requires more radiation resistant, sensitive, and fast scintillating materials. 

In this regard, we have investigated radiation hardness and timing characteristics of plastic scintillators such as Polyethylene Naphthalate (PEN), Polyethylene Terephthalate (PET), Eljen brand EJ-260 (EJN) and overdoped EJ-260 (EJ2P), and a lab produced elostemer scintillator (SX). However, the scope of this study is limited to the comparison of PEN and PET samples only in terms of radiation hardness and timing characteristics.  

PEN and PET emit blue light with peak emission wavelengths of 450 nm and 350 nm and they produce 10,500 and 2,200 photons per MeV, respectively \cite{Nakamura}. PEN scintillators were previously used in radioactive dosimeters in nuclear experiments and medical physics \cite{Fluhs} but none of them have been used in particle physics experiments. We would like to introduce PEN and PET to high energy particle experiments. 

\section{Experimental Setup}
Two samples of each material were irradiated up to 1.4 and 14 Mrad using gamma radiation from a $^{137}$Cs source at the University of Iowa RadCore Facility. Their radiation hardness in terms of the light yield was measured afterwards. The sample size of the PEN and PET tiles are 10 cm x 10 cm x 0.1 cm and 10 cm x 10 cm x 0.2 cm, respectively. They were uniformly exposed to the radiation source.  

\begin{figure}[h!]
\centering 
\includegraphics[width=0.45\textwidth]{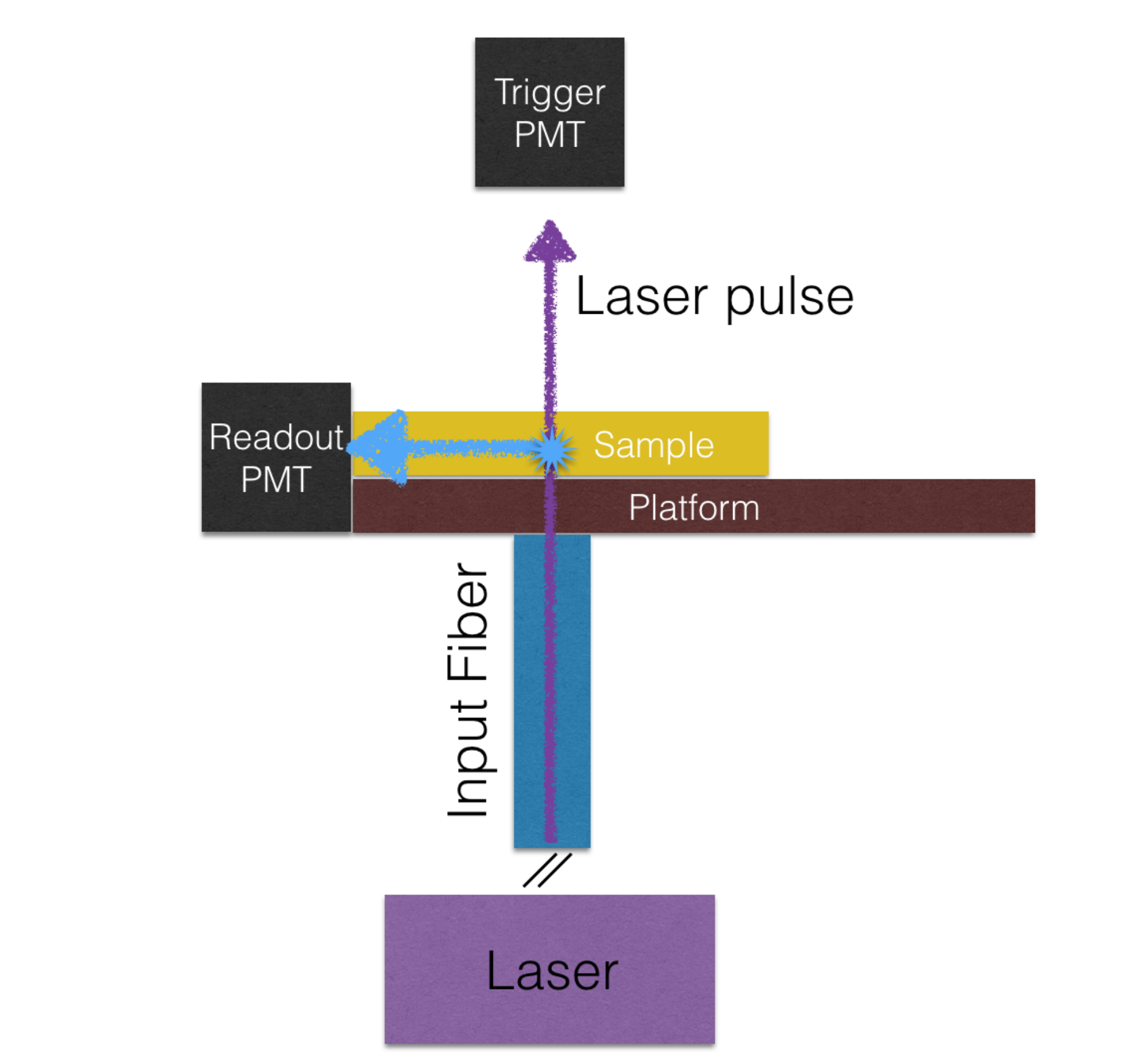}
\caption{Experimental setup for the light yield measurement} 
\label{ExperimentalSetup}
\end{figure}

Before and after the irradiation, the light yield of the samples were measured by using the experimental setup in Fig. \ref{ExperimentalSetup}. We used a control sample to calibrate the experimental setup before every measurement. Each sample was kept and tested in a temperature and humidity controlled room at 22.5 C$^{\circ}$.  

In the experimental setup, we pulsed a 337 nm nitrogen laser with a 3 ns pulse width perpendicular to the surface of the sample. The scintillation light was read out from the side with a directly coupled Hamamatsu R7600 PMT, triggered by a Hamamatsu R7525 PMT on the top of the setup. 
 
Their timing characteristics were also studied with beam test at Fermilab Test Beam Facility (FTBF) in a different experimental setup.   

 \section{Measurement Results} 
\begin{figure}[h!]
\centering 
\includegraphics[width=0.75\textwidth]{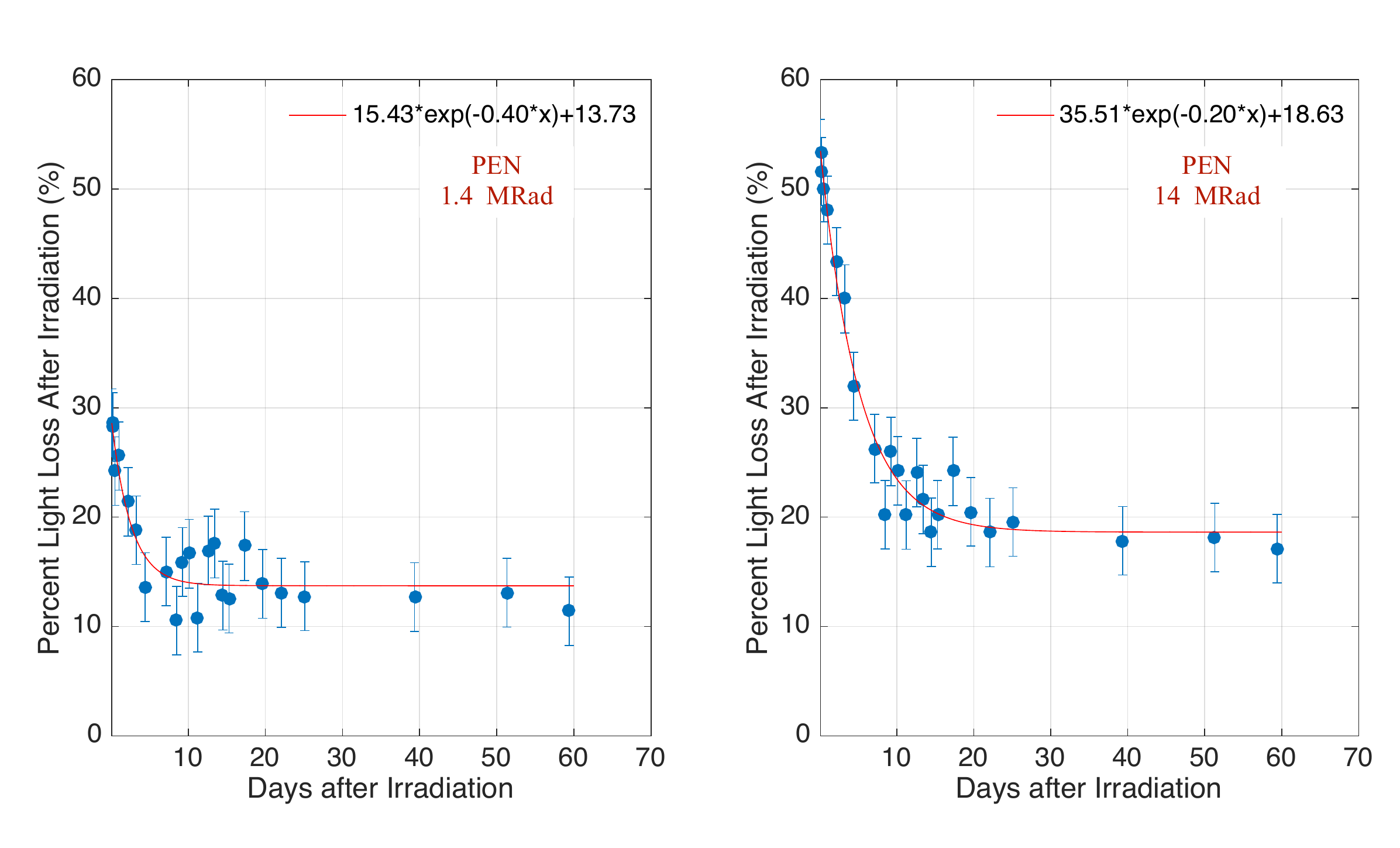}
\includegraphics[width=0.75\textwidth]{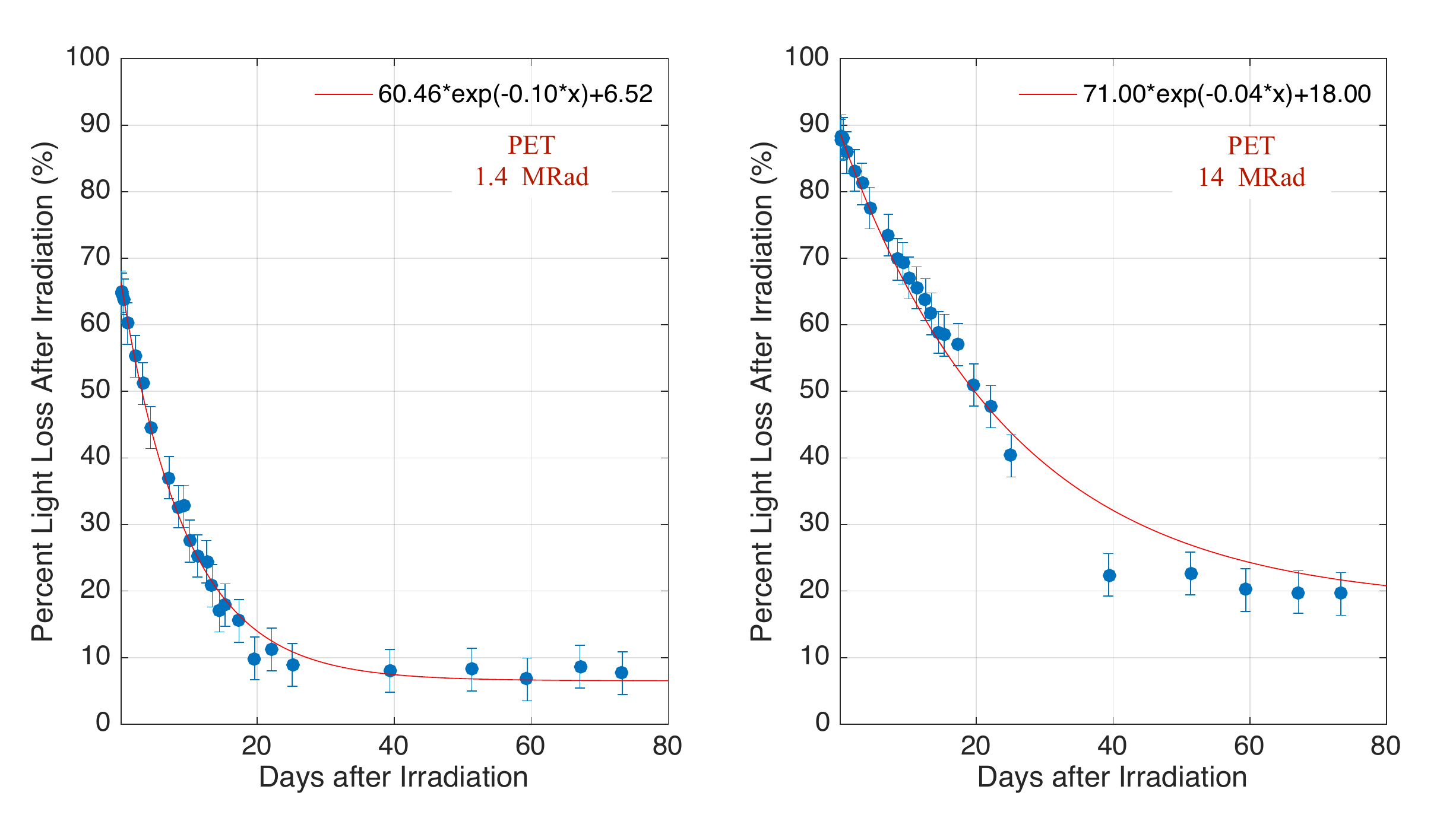}
\caption{PEN (top left at 1.4 Mrad and top right at 14 Mrad) and PET (bottom left at 1.4 Mrad and bottom right at 14 Mrad) light yield results over 50 days after irradiation.}
\label{Results}
\end{figure}

Figure \ref{Results} shows the percent light loss after irradiation as a function of time for PEN at 1.4 Mrad (top left) and at 14 Mrad (top right), and PET at 1.4 Mrad (bottom left) and at 14 Mrad (bottom right) \cite{Wetzel2016}.

Table \ref{PENPETSummary} shows initial light yield, recovered light yield and recovery time of PEN and PET, which are concluded from Fig. \ref{Results}. Here, initial light yield is the measured light yield compared to undamaged sample immediately after the exposure. After some time the tile recovers from the radiation damage and the light yield reaches to a maximum value (a plateau), which gives the recovered light yield and the recovery time is the time takes a sample to reach that plateau value.  

\begin{table}[h!]
\caption{Summary of the PEN and PET light yield results}
\label{PENPETSummary}
\begin{center}
\begin{tabular}{lcccccc}
\toprule
\multirow{2}{*}{} & 
\multicolumn{2}{c}{Initial Light Yield (\%)} &
\multicolumn{2}{c}{Recovered Light Yield (\%)} &
\multicolumn{2}{c}{Recovery Time (days)} \\
      & {1.4 Mrad} & {14 Mrad} & {1.4 Mrad} & {14 Mrad} & {1.4 Mrad} & {14 Mrad} \\
      \midrule
    PEN & 71.4 & 46.7 & 85.9 & 79.5 & 5 & 9 \\
    PET & 35.0 & 12.2 & 93.5 & 80.0 & 22 & 60 \\
    \hline
\end{tabular}
\end{center}
\end{table}

Initially, PEN was damaged 28.6\% at 1.4 Mrad and 53.3\% at 14 Mrad and PET was damaged 65\% at 1.4 Mrad and 87.8\% at 14 Mrad. This shows that PET was damaged more than PEN initially. After the recovery, the permanant damage on PET is 14.1\% at 1.4 Mrad and 20.5\% at 14 Mrad, and the permanent damage on PET is 6.5\% at 1.4 Mrad and 20\% at 14 Mrad. The permanent damage is almost same for both scintillators at 14 Mrad and PEN needs much shorter time than PET to recover.  

Also, the timing characteristics of PEN and PET were studied at Fermilab Test Beam Facility (FTBF) with beam test. Signal timing, peak to 1/e values of PEN and PET were calculated and compared to SCSN-81, which is a known scintillator at Compact Muon Solenoid (CMS) experiment at Large Hadron Collider (LHC). Peak to 1/e values for PEN, PET and SCSN-81 are respectively 27.12 ns, 6.884 ns and 10.56 ns \cite{Tiras2015}. 

\section{Conclusion}
Polyethylene Naphthalate (PEN) and Polyethylene Teraphthalate (PET) samples were irradiated up to 1.4 and 14 Mrad with a $^{137}$Cs gamma radiation source. PET was damaged initially more than PEN but the permanent damage on them is almost same. PEN was recovered more quickly than PET. However, PET has much faster response time than PEN. With all these results, we aim to investigate a blended sample of PEN and PET.


\begin{thebibliography}{99}
\bibitem{Nakamura}
H. Nakamura et al.,
\newblock {Evidence of deep-blue photon emission at high efficiency by common plastic},  
\newblock {\textit{Europhys. Lett. 95 (2011) 22001}, 2011}. 

\bibitem{Fluhs}
D. Fluhs et al., 
\newblock{Polyethylene Naphthalate Scintillator: A Novel Detector for the Dosimetry of Radioactive Ophthalmic Applicators} 
\newblock{\textit{Ocular Oncology and Pathology}, v.2(1), September 2015.} 

\bibitem{Wetzel2016}
J. Wetzel et al., 
\newblock {Radiation Damage and Recovery Properties of Common Plastics PEN (Polyethylene Naphthalate) and PET (Polyethylene Terephthalate) Using a 137Cs Gamma Ray Source Up To 1.4 Mrad and 14 Mrad},  
\newblock {\textit{Journal of Instrumentation},V11-P08023, August 2016}. 

\bibitem{Tiras2015}
E. Tiras, 
\newblock {Radiation Hard \& High Light Yield Scintillator Search for CMS Phase II Upgrade},  
{arXiv:1510.08572}, 
\newblock {\textit{Proceedings, Meeting of the APS DPF}, Ann Arbor, Michigan, USA, 4-8 Aug 2015}. 



\end{thebibliography}
\end{document}